\documentclass[aps,prd,eqsecnum,preprint,nofootinbib]{revtex4}
\usepackage[dvipdfmx]{color,graphicx}
\usepackage{color}
\usepackage{graphicx}
\usepackage{dcolumn}
\usepackage{bm} 
\usepackage{amssymb}
\usepackage{amsfonts}
\usepackage{amsmath}
\usepackage{color}
\allowdisplaybreaks[1]

\definecolor{DarkBlue}{rgb}{0,0,0.7} 

\definecolor{DarkRed}{rgb}{0.65,0,0} 

\definecolor{DarkGreen}{rgb}{0,0.6,0}

\begin{document}
\author{Taishi~Ikeda}\email{ikeda@gravity.phys.nagoya-u.ac.jp}
\affiliation{
Departure of Physics,~~Graduate School of Science,~~Nagoya University,~~Nagoya~~464-6602,~~Japan}

\author{Chul-Moon~Yoo}\email{yoo@gravity.phys.nagoya-u.ac.jp}
\affiliation{
Departure of Physics,~~Graduate School of Science,~~Nagoya University,~~Nagoya~~464-6602,~~Japan}

\title{Critical behavior of 
spherically symmetric domain wall collapse}

\begin{abstract}
Critical collapse of 
a spherically symmetric domain wall is investigated. 
The domain wall is made of
a minimally coupled scalar field with 
a double well potential.
We consider 
a sequence of the initial data which describe a momentarily static 
domain wall characterized by its initial radius. 
The time evolution is performed by a full general relativistic 
numerical code for spherically symmetric systems.
In this paper, 
we use the
maximal slice gauge condition, 
in which spacelike time slices 
may penetrate the black hole horizon
differently from other widely used procedures. 
In this paper, we consider two specific shapes of the double well potential,
and observe the Type II critical behavior in both cases.
The mass scaling, sub-critical curvature scaling, 
and those fine structures are confirmed.
The index of 
the scaling behavior
agrees with the massless scalar case.
\end{abstract}

\maketitle

\section{Introduction}
Solutions for the Einstein equations and their 
non-linear dynamics have rich structure and behavior, and 
they have been extensively investigated in many aspects for a century. 
After numerical techniques
were established, 
research on the non-linear dynamics has been developed a lot, 
and still continues to attract much attention up to the present date. 
The critical collapse, which was firstly observed by Choptuik~\cite{Choptuik:1992jv} 
for the spherically symmetric massless scalar system with numerical simulation, 
is one of the most interesting discoveries in non-linear gravitational dynamics.
After the Choptuik's discovery, 
many researchers have discovered the critical collapse in
several systems such as
axisymmetric gravitational wave collapse~\cite{Abrahams:1993wa,Sorkin:2010tm}, 
spherically symmetric gravitational collapse of
a perfect fluid~\cite{Evans:1994pj,Maison:1995cc} and 
collapse of collisionless matter particles~\cite{Olabarrieta:2001wy}.

Let us consider a collapsing system whose initial data are characterized by 
a parameter $p$, and suppose that 
a black hole is finally formed for 
the solutions with $p>p_{\ast}$(supercritical), 
while it is not the case for $p<p_{\ast}$(subcritical). 
The critical behavior can be observed near the threshold of 
the black hole formation $p\sim p_{\ast}$.
It has some characteristic features of 
the intermediate state of the gravitational collapse and
the behavior of the black hole mass in the supercritical region. 
The intermediate state of the gravitational collapse around 
the threshold does not depend on how to parametrize the initial data,
and it is called a critical solution.

Here, we focus on the so-called Type II critical collapse
associated with a discrete self-similar critical solution (see  e.g., 
a review paper~\cite{Gundlach:2007gc} for general cases). 
In this case,
the black hole mass in the supercritical region obeys 
the scaling law with a periodic fine structure as follows:
$\ln M_{\mbox{\footnotesize{BH}}} = \nu\ln |p-p_{\ast}| + c + f(\ln |p-p_{\ast}|)$, 
where $c$ is a constant, 
and $f(x)$ is a periodic function 
satisfying $f(x+\varpi)=f(x)$.
The index $\nu$ and period $\varpi$ 
take universal values irrespective of 
the parametrization of the initial data. 
An example of this type is
the spherically symmetric massless scalar 
system~\cite{Choptuik:1992jv,Hod:1996az}, 
and we obtain 
$\nu\sim0.377$ and $\varpi\sim 4.6$ in this case. 

Scaling lows can be also observed 
not only in the super-critical region but also in the 
subcritical region for some physical quantities. 
Garfinkle reported the following scaling laws of the maximum values 
of the curvature invariants at the origin of a spherical system:
$|R|_{\mbox{\footnotesize{max}}}\propto |p-p_{\ast}|^{-2\nu}$ and 
$|R_{\mu\nu}R^{\mu\nu}|_{\mbox{\footnotesize{max}}}\propto |p-p_{\ast}|^{-4\nu}$, 
where $\nu$ is the index of the black hole mass scaling~\cite{Garfinkle:1998va}.
It has been revealed that such rich phenomena of the critical collapse 
is related 
to the structure of the phase space~\cite{Koike:1995jm,Gundlach:1996eg}. 
Other related works are 
summarized in a review paper~\cite{Gundlach:2007gc}. 

Finding universal phenomena in complicated non-linear dynamics 
is important and useful to understand characteristic features of the theory. 
The critical collapse may reflect a universal feature of gravitational theory. 
Investigation into the critical collapse may provide
not only deep understanding of gravitational theory but also 
astrophysical implication. 
For instance, 
the threshold for the black hole formation 
is particularly important for the primordial black holes. 
Some of the primordial black holes would have been formed in the early universe 
as a result of near critical collapse. 
Therefore, the scaling behavior would be important and useful to 
predict the mass spectrum of the number density of 
the primordial black holes~\cite{Niemeyer:1997mt}.

Choptuik discussed
the critical collapse of the scalar field 
with a polynomial potential~\cite{Choptuik:1994ym}, 
and concluded that the polynomial potential is irrelevant under the critical collapse,
because the kinetic term of the scalar field dominates under the discrete self-similar spacetime. 
That is, if the critical behavior is realized 
for the scalar field with a polynomial potential
with a discrete self-similar spacetime 
as the intermediate state, 
it is expected that the scaling law 
is identical to the massless case.
However, in general, 
it is nontrivial 
if the critical collapse with a discrete self-similar spacetime 
is actually realized or not with a polynomial potential.
In fact, as is reported
in Refs.~\cite{Brady:1997fj,Okawa:2013jba}, 
the critical collapse of the massive scalar system 
can be different from the massless case depending on the 
relative value of 
the mass of the scalar field 
to the physical 
scale characterizing the initial data. 

In this paper, 
we simulate spherically symmetric domain wall collapse of
a scalar field with  
a double well potential. 
Dynamics of a spherically symmetric domain wall 
has been investigated in Ref.~\cite{Widrow:1989vj} 
on the flat background. 
It has been shown that a black hole can be formed 
by gravitational collapse of 
the domain wall in Ref.~\cite{Garfinkle:2012hx}. 
Astrophysical implication of the primordial black holes 
from domain wall collapse is discussed in Ref.~\cite{Rubin:2000dq}.
Recently, it was reported that 
mass scaling appears in the domain wall collapse and 
the index of the scaling agrees 
with the index 
in the massless scalar system~\cite{Clough:2016jmh}. 
While, the fine structure 
has not been confirmed yet. 
As is noted above, 
depending on the value of the parameter for the potential, 
there is a possibility to obtain different behavior 
from the Type II critical behavior. 
Therefore, in this paper, 
we adopt different parameter sets from those in Ref.~\cite{Clough:2016jmh},
and discuss whether the mass scaling and the fine structure appear. 

One remarkable feature of our work 
is in the gauge condition for time slices.
In many papers about the critical collapse, 
the areal polar gauge condition or null coordinates are used. 
In the areal polar gauge,
the black hole horizon is identified by the vanishing lapse function.
Therefore, time slices cannot penetrate the black hole horizon. 
In the case of null coordinates, 
the apparent horizon is foliated by null surfaces 
starting from the outer-boundary in the asymptotic region. 
While, in our case, we get a foliation of spatial sections 
of the apparent horizon. 
There are a few papers in which 
a similar gauge condition to ours 
is used~\cite{Olabarrieta:2001wy,Okawa:2013jba,Cao:2016tvh}.
In these papers, and also in this paper, 
the black hole mass is defined 
by the half of the areal radius of the apparent horizon 
at the moment of the apparent horizon formation.
Obviously, 
the definition of the black hole mass depends on 
the time slice.
However, in spite of this fact,
the critical behavior appears,
as is reported in Refs.~\cite{Olabarrieta:2001wy,Okawa:2013jba,Cao:2016tvh}.
It should be noted that, 
the areal radius of the apparent horizon with 
a spatial section is, of course, different from the 
areal radius of the event horizon near the null infinity. 
Therefore, scaling behaviors of these two horizons should be 
independently discussed and compared with each other.

This paper is organized as follows. 
In the next section, 
we explain our formulation and numerical schemes.
Initial data for the numerical simulation is 
shown in
section~\ref{initial data}. 
In section~\ref{result}, 
we provide our numerical results,
and discuss the mass scaling, fine structure and curvature scaling.
Section~\ref{summary} is devoted to a summary and discussion.

Throughout this paper, 
we use the geometrized units in which 
the speed of light and Newton's gravitational constant are one, 
respectively. 

\section{Formulation and numerical schemes}\label{Set up}
\subsection{Formulation}
We consider the Einstein-scalar system with 
a double well potential whose Lagrangian is given by
\begin{equation}
S=\int d^{4}x\sqrt{-g}\left\{ \frac{R}{16\pi}-\frac{1}{2}\nabla_{\mu}\Phi\nabla^{\mu}\Phi-V(\Phi)\right\}, 
\end{equation}
where $g_{\mu\nu}$ is the metric tensor, 
$R$ is the Ricci scalar and $\Phi$ is a scalar field.
$V(\Phi)$ is the double well potential given by
\begin{equation}
V(\Phi)=\frac{\lambda}{24}(\Phi^{2}-\sigma^{2})^{2},
\end{equation}
where $\lambda$ and $\sigma$ are 
constant parameters. 
We focus on
spherically symmetric spacetimes, whose 
line element is described by
\begin{eqnarray}
ds^{2}&=&-\alpha^{2}(t,r)dt^{2}+\psi^{4}(t,r)\left\{\gamma (t,r)^{-2}(dr+r\beta(t,r)dt)^{2}+\gamma(t,r) r^{2}d\Omega^2\right\}, \label{metric}
\end{eqnarray}
where 
$d\Omega^{2}$ is the solid angle element, and $\psi$, $\alpha$, $\beta$ and 
$\gamma$ are independent functions of $t$ and $r$. 
Under this ansatz, 
the 3-metric $\gamma_{ij}$ is expressed as follows:
\begin{equation}
\gamma_{ij}=\psi^{4}\mbox{diag}(\gamma^{-2},\gamma r^{2},\gamma r^{2}\sin^{2}\theta).
\end{equation}
The extrinsic curvature $K_{ij}$ of each time slice 
can be expressed by using the two independent components 
$K\equiv\gamma^{ij}K_{ij}$ and 
$A\equiv(K_{\theta\theta}-\frac{1}{3}K\gamma_{\theta\theta})/(\psi^{4}r^{2})$.

Substituting the metric form (\ref{metric}) 
into the Einstein equations 
and the equation of motion for the scalar field,
we get the following time evolution equations: 
\begin{eqnarray}
\displaystyle (\partial_{t}-r\beta\partial_{r})\psi
&\displaystyle =
&\displaystyle \frac{1}{6}\psi(3\beta+r\beta^{\prime}-\alpha K),\label{time evolution psi}\\
\displaystyle (\partial_{t}-r\beta\partial_{r})K
&\displaystyle =
&\displaystyle \alpha\left\{\frac{1}{3}K^{2}+6\frac{A^{2}}{\gamma^{2}}+8\pi\Pi^{2}-8\pi V(\Phi)\right\}-\psi^{-4}\gamma^{2}\left\{\Delta\alpha+2\alpha^{\prime}(\frac{\psi^{\prime}}{\psi}+\frac{\gamma^{\prime}}{\gamma})\right\},\\
\displaystyle (\partial_{t}-r\beta\partial_{r})\gamma
&\displaystyle =
&\displaystyle -2\alpha A-\frac{2}{3}r\gamma\beta^{\prime},\label{time evolution gamma}\\
\displaystyle (\partial_{t}-r\beta\partial_{r})A
&\displaystyle =
&\displaystyle \alpha K A -2\alpha\frac{A^{2}}{\gamma}-\frac{2}{3}rA\beta^{\prime}+\psi^{-4}\left\{-\frac{1}{6}\gamma^{3}(\Delta\alpha-3\alpha^{\prime\prime})-\frac{1}{3}\alpha\gamma^{3}(\frac{\Delta\psi}{\psi}-3\frac{\psi^{\prime\prime}}{\psi})\right.\nonumber\\
&&\displaystyle -\frac{1}{6}\alpha(1+\gamma)\Delta\gamma+\frac{1}{6}\alpha(1+\gamma+\gamma^{2})\gamma^{\prime\prime}-\frac{1}{3}\alpha(1+\gamma+\gamma^{2})(-\frac{\gamma-1}{r^{2}}+\frac{\gamma^{\prime}}{r})+\frac{1}{6}\alpha^{\prime}\gamma^{2}\gamma^{\prime}\nonumber\\
&&\displaystyle \left.-\frac{4}{3}\alpha^{\prime}\gamma^{3}\frac{\psi^{\prime}}{\psi}+\frac{1}{3}\alpha\gamma^{2}\gamma^{\prime}\frac{\psi^{\prime}}{\psi}-2\alpha\gamma^{3}\frac{\psi^{\prime 2}}{\psi^{2}}+\frac{8}{3}\pi\alpha\gamma^{3}\Phi^{\prime 2}\right\},\label{time evolution A}\\
\displaystyle (\partial_{t}-r\beta\partial_{r})\Phi
&\displaystyle =
&\displaystyle-\alpha\Pi\label{time evolution Phi},\\
\displaystyle (\partial_{t}-r\beta\partial_{r})\Pi
&\displaystyle =
&\displaystyle\alpha\Pi K-\psi^{-4}\alpha\gamma^{2}\left\{\Delta\Phi+2\Phi^{\prime}(\frac{\gamma^{\prime}}{\gamma}+\frac{\psi^{\prime}}{\psi}+\frac{\alpha^{\prime}}{2\alpha})\right\}+\alpha V^{\prime}(\Phi),\label{time evolution Pi}
\end{eqnarray}
where $\Pi$ is the conjugate momentum
of $\Phi$.
The 
Hamiltonian constraint
and the momentum constraint
can be written as
\begin{eqnarray}
\displaystyle \frac{\Delta\psi}{\psi}
+\frac{1}{8}(5\frac{\Delta\gamma}{\gamma}-3\frac{\gamma^{\prime\prime}}{\gamma})
+\pi\Phi^{\prime2}
+2\pi\gamma^{-2}\psi^{4}V(\Phi)
+\frac{(\gamma^{2}+\gamma+1)(\gamma-1)}{4\gamma^{3}r^{2}}\nonumber\\
\displaystyle +\frac{\gamma^{\prime}}{\gamma}(2\frac{\psi^{\prime}}{\psi}+\frac{3}{16}\frac{\gamma^{\prime}}{\gamma})
+\frac{\psi^{4}}{\gamma^{2}}(\frac{3A^{2}}{4\gamma^{2}}+\pi\Pi^{2}-\frac{1}{12}K^{2})&=&0\label{Ham con},\\
\displaystyle 
A^{\prime}+\frac{\gamma}{3}K^{\prime}+4\pi\gamma\Pi\Phi^{\prime}+\frac{3A}{r}+\frac{A\gamma^{\prime}}{2\gamma}+6A\frac{\psi^{\prime}}{\psi}&=&0
\label{Mom con}.
\end{eqnarray}

The boundary condition for the variables at the origin is
imposed by the Neumann boundary condition 
in order to guarantee regularities. 
In addition, 
we require that
$\gamma$ and $A$ satisfy the local flatness condition:
\begin{equation}\label{local flatness condition}
\gamma(t,r=0)=1,~A(t,r=0)=0. 
\end{equation}
The Neumann boundary condition
and the local flatness condition regularize the $1/r$ and $1/r^{2}$ terms
in Eq.~\eqref{time evolution A} and constraint equations. 
It can be straightforwardly checked that,
if the condition \eqref{local flatness condition} 
is satisfied on the initial hypersurface and 
the time evolution equations are 
exactly satisfied, 
the condition \eqref{local flatness condition} 
is kept for every time step.
However, 
when the time evolution equations are numerically solved,
the local flatness condition can be violated due to numerical errors. 
When 
the Neumann boundary condition for $A$ and $\gamma$ is imposed,
the local flatness condition cannot be explicitly enforced. 
In consequence, numerical instability 
may be generated by $1/r$ and $1/r^2$ terms in the original 
evolution equation \eqref{time evolution A} for $A$.
In order to avoid the numerical instability, 
instead of Eq.~\eqref{time evolution A}, 
we solve the momentum constraint (\ref{Mom con}) for $A$ 
at each time step. 
At the origin, the equation (\ref{Mom con}) is evaluated by using
the Neumann boundary condition and 
the local flatness condition 
\eqref{local flatness condition}. 

While the appearance of $1/r$ and $1/r^2$ terms is 
avoided in the evolution equation by the prescription stated above, 
we still have those terms in the Hamiltonian constraint \eqref{Ham con}. 
Due to these terms, we suffer from the large cancellation error in 
evaluation of the constraint violation near the origin, 
and we cannot use the Hamiltonian constraint \eqref{Ham con} to 
check the accuracy of the numerical computation. 
In order to overcome this difficulty, 
we introduce
the auxiliary field $\Gamma(t,r)$ defined by
\begin{equation}
\Gamma\equiv \gamma^{\prime}+\frac{3}{r}(\gamma-1).
\label{def Gam}
\end{equation}
Then, the constraint equation (\ref{Ham con}) becomes
\begin{eqnarray}
\displaystyle \frac{\Delta\psi}{\psi}
+\frac{1}{8}(5\frac{\Delta\gamma}{\gamma}-3\frac{\gamma^{\prime\prime}}{\gamma})
+\pi\Phi^{\prime2}
+2\pi\gamma^{-2}\psi^{4}V(\Phi)
+\frac{\gamma^{2}+\gamma+1}{8\gamma^{3}}(\Delta\gamma-\frac{1}{3}\gamma^{\prime\prime}-\frac{2}{3}\Gamma^{\prime})\nonumber\\
\displaystyle +\frac{\gamma^{\prime}}{\gamma}(2\frac{\psi^{\prime}}{\psi}+\frac{3}{16}\frac{\gamma^{\prime}}{\gamma})
+\frac{\psi^{4}}{\gamma^{2}}(\frac{3A^{2}}{4\gamma^{2}}+\pi\Pi^{2}-\frac{1}{12}K^{2})&=&0.\hspace{1cm}
\label{Ham con new}
\end{eqnarray}
From the time evolution equations, 
we can derive the evolution equation for $\Gamma$ as follows:
\begin{eqnarray}
\displaystyle (\partial_{t}-r\beta\partial_{r})\Gamma
&\displaystyle =\alpha\frac{\gamma^{\prime}}{\gamma}A+8\pi\alpha\gamma\Pi\Phi^{\prime}+12\alpha A\frac{\psi^{\prime}}{\psi}+\frac{2}{3}\alpha\gamma K^{\prime}-2\alpha^{\prime}A-\frac{8}{3}\gamma\beta^{\prime}\nonumber\\
&\displaystyle +\frac{1}{3}r\gamma^{\prime}\beta^{\prime}-\frac{2}{3}r\gamma\beta^{\prime\prime}+\beta\Gamma.\label{time evolution Gamma}
\end{eqnarray}
We check the violation of Eqs.~\eqref{def Gam} and \eqref{Ham con new} throughout 
the time evolution as measures of numerical accuracy. 

The evolution equation for $A$, which we do not solve in this work, 
can be also derived from the evolution equation as follows:
\begin{eqnarray}
\displaystyle (\partial_{t}-r\beta\partial_{r})A
&\displaystyle =\alpha K A -2\alpha\frac{A^{2}}{\gamma}-\frac{2}{3}rA\beta^{\prime}+\psi^{-4}\left\{-\frac{1}{6}\gamma^{3}(\Delta\alpha-3\alpha^{\prime\prime})-\frac{1}{3}\alpha\gamma^{3}(\frac{\Delta\psi}{\psi}-3\frac{\psi^{\prime\prime}}{\psi})\right.\nonumber\\
&\displaystyle -\frac{1}{6}\alpha(1+\gamma)\Delta\gamma+\frac{1}{18}\alpha(1+\gamma+\gamma^{2})\gamma^{\prime\prime}+\frac{1}{9}\alpha(1+\gamma+\gamma^{2})\Gamma^{\prime}+\frac{1}{6}\alpha^{\prime}\gamma^{2}\gamma^{\prime}\nonumber\\
&\displaystyle \left.-\frac{4}{3}\alpha^{\prime}\gamma^{3}\frac{\psi^{\prime}}{\psi}+\frac{1}{3}\alpha\gamma^{2}\gamma^{\prime}\frac{\psi^{\prime}}{\psi}-2\alpha\gamma^{3}\frac{\psi^{\prime 2}}{\psi^{2}}+\frac{8}{3}\pi\alpha\gamma^{3}\Phi^{\prime 2}\right\}. 
\label{time evolution A new}
\end{eqnarray}
We note that, in this scheme, 
$1/r$ and $1/r^{2}$ terms are also removed from 
the time evolution equation \eqref{time evolution A new}(see also 
Ref.~\cite{Alcubierre:2004gn}). 
It is also worth to note that 
this procedure is quite similar to the 
so-called BSSN scheme in numerical relativity~\cite{Shibata:1995we,Baumgarte:1998te}.

\subsection{Gauge condition}
For the shift vector, 
we simply set
\begin{equation}
\beta(t,r)=0. 
\end{equation}
For the lapse function,
we use the
maximal slice condition $K=0$. 
By this condition, 
the time evolution equation for $K$ 
gives the elliptic partial differential equation for $\alpha(t,r)$:
\begin{eqnarray}
\Delta\alpha+2\alpha^{\prime}(\frac{\psi^{\prime}}{\psi}+\frac{\gamma^{\prime}}{\gamma})=\alpha\psi^{4}\gamma^{-2}\left\{6\frac{A^{2}}{\gamma^{2}}+8\pi\Pi^{2}-8\pi V(\Phi)\right\}.\label{maximum slice}
\end{eqnarray}
Because the solution for this equation has the ambiguity of 
a constant factor, 
we normalize it so that the lapse function 
may be unity at far boundary. 

The lapse 
$\alpha$ is 
a non-trivial function 
of the radial coordinate, 
and it may have the value larger than unity for some cases.
In such cases, we need to care about
the Courant-Friedrichs-Lewy condition (CFL condition). 
In order to 
appropriately impose the CFL condition for the physical time scale,
we normalized the time step interval $\Delta t$ as 
$\alpha_{\mbox{\footnotesize{max}}}\Delta t=0.75\Delta$,
where $\Delta$ is the grid interval of the radial coordinate and
$\alpha_{\mbox{\footnotesize{max}}}$ is the maximum value of the lapse function 
at each time step.

\subsection{Numerical scheme}
We implemented the equations (\ref{time evolution gamma}), (\ref{time evolution Phi}), 
(\ref{time evolution Pi}), (\ref{Mom con}) and (\ref{maximum slice})
in a numerical code.
Because we use the maximal slice condition,
the time derivative of $\psi$ vanishes,
and we do not need to solve the Eq.~(\ref{time evolution psi}).
The integration in time is 
performed by
the iterative Crank-Nicolson scheme\cite{Teukolsky:1999rm},
and 
spatial derivatives 
are evaluated by using a
4th order finite difference method
except for $\gamma$ in $r<0.01/\mu$, where
spatial derivatives of $\gamma$ is evaluated 
by using the 2nd order central difference method. 
The laplacian terms in each equation at the origin 
are evaluated by using the CARTOON method in $r<0.01/\mu$~\cite{Alcubierre:1999ab}.

\section{initial data}\label{initial data}

Let us consider a momentarily static domain wall 
based on the isotropic coordinate $\tilde{r}$, 
for which a line element $dh^2$ on the initial hyper-surface can be expressed as 
\begin{equation}
dh^2=\tilde \psi^4 \left(d\tilde r^2+\tilde r^2d\Omega^2\right). 
\label{ini met}
\end{equation}
The relation between the coordinate $r$ and $\tilde r$ will be 
given later.
From the momentarily static condition, 
$K$, $A$ and $\Pi$ 
are assumed to be zero at $t=0$. 
Consequently, 
the momentum constraint is trivially satisfied. 

\begin{figure}
\begin{tabular}{cc}
\includegraphics[scale=0.65,clip]{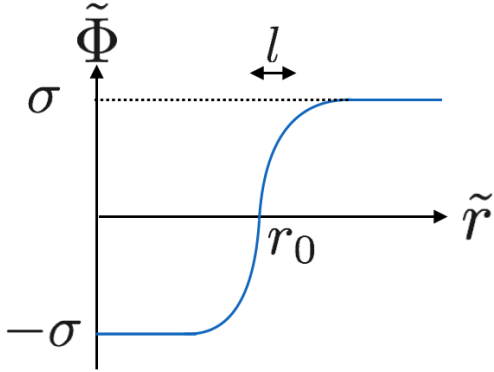}
\end{tabular}
\caption{\label{initial_data.eps}
The profile of the scalar field 
for the initial data.
}
\end{figure}

The initial scalar field profile is assumed to be the following form:
\begin{eqnarray}
\displaystyle \Phi|_{t=0}=\sigma\tanh\left(\frac{\tilde r-r_{0}}{l}\right)
+\sigma\left\{-1-\tanh\left(\frac{\tilde r-r_{0}}{l}\right)\right\}
\exp\left\{-\left(\frac{\tilde r}{l}\right)^4\right\},
\end{eqnarray}
where $r_{0}$ is the initial radius of the domain wall and $l$ is the width of the domain wall.
The width $l$ is given by
the value of the planar domain wall solution as
$l=\frac{2}{\sigma}\sqrt{\frac{3}{\lambda}}\equiv\sqrt{\frac{2}{\mu^{2}}}$ (see Fig\ref{initial_data.eps}). 
The first term locally describes the planar domain wall solution 
for large $\tilde r$.
While the second term is introduced to regularize 
the scalar filed at the origin.

Because it is expected that curvature 
values for near critical cases are very large, 
high numerical resolution is needed especially in the central region
to resolve the critical behavior. 
In order to realize the finer resolution near the origin, 
we use the coordinate $r$ which is related to 
the isotropic coordinate $\tilde r$ as follows:
\begin{equation}\label{inhomogeneous grid}
r=\left\{
\begin{array}{cc}
\tilde r&(0<\tilde r<R), \\
\tilde r+\left(\frac{\tilde r-R}{w}\right)^{\rho}&(R<\tilde r),
\end{array}\right.
\end{equation}
where $R$, $w$ and $\rho$ are 
constant parameters.
Performing the coordinate transformation 
from $\tilde r$ to $r$, $\gamma$ can be read off 
from the spatial part of the metric \eqref{metric}. 
$\Gamma$ can be also calculated by the definition \eqref{def Gam}. 
Then,
solving the Hamiltonian constraint equation \eqref{Ham con} for
$\psi$,
we can obtain the initial data set. 

It is worthy to note that, by virtue of the new coordinate $r$, 
the physical distance to the outer numerical boundary 
can be so large that the scalar waves cannot reach the outer boundary.
Therefore, 
during the time evolution, 
the asymptotically flat boundary condition 
can be easily implemented for the radial coordinate $r$.


\section{results}\label{result}
\subsection{Setup}
Hereafter we express all the variables in the unit of $\mu$.
The radius of the domain wall $r_{0}$ is 
the only parameter which characterizes the initial data of this system. 
We have another parameter $\lambda$ to specify the potential shape.
In Ref.~\cite{Clough:2016jmh}, 
the cases $\lambda=30000\mu^2$ and $\lambda=60000\mu^2$ are investigated. 
We consider smaller values of $\lambda/\mu^2$ and higher 
potential barrier between the potential minima 
than the cases in Ref.~\cite{Clough:2016jmh}, 
so that we can observe characteristic behavior due to existence of the two 
potential minima. 
In this paper, we investigate the cases $\lambda=1000\mu^2$ and $\lambda=2000\mu^2$.
As for the
parameters in the coordinate transformation \eqref{inhomogeneous grid}, 
we use the following two parameter sets: 
\begin{eqnarray}
{\rm Param-1}&:&~~R=0.1/\mu,~w=0.05/\mu,~\rho=6~{\rm for}~0<r<0.2, \\
{\rm Param-2}&:&~~R=0.03/\mu,~w=0.01/\mu,~\rho=6~{\rm for}~0<r<0.05.
\end{eqnarray}
In both cases, the maximum value of the areal radius is about $60/\mu$.

\subsection{Convergence test}
\begin{figure}
\begin{center}
\includegraphics[scale=1]{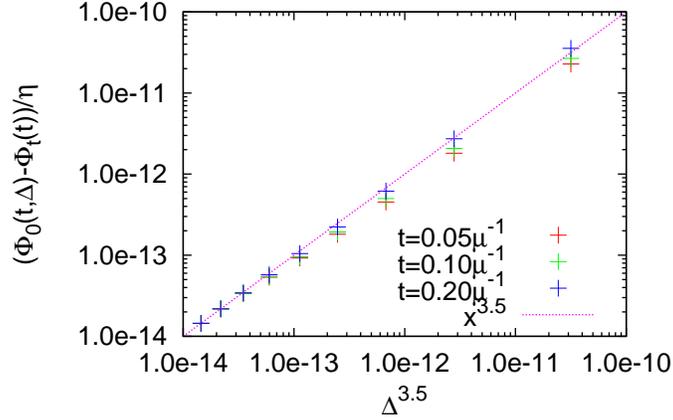}
\caption{
The convergence of $\Phi_0(t,\Delta)$ for each time step. 
}
\label{convergence}
\end{center}
\end{figure}

Before showing main results, in this subsection, 
we present the result of a convergence test for our numerical code
by using simulation with $r_0=0$ for each resolution. 
We show the result for $\lambda=2000\mu^2$. 
For simplicity, we focus on the value $\Phi_0(t,\Delta):=\Phi(t,r=0;\Delta)$, 
where we have explicitly written the dependence on the 
grid interval $\Delta$. 
If our numerical code obeys the $n$-th order convergence, 
we obtain
\begin{equation}\label{convergence fitting function}
\Phi_0(t,\Delta)=\Phi_{\rm t}(t)+\eta\Delta^{n},
\end{equation}
where $\Phi_{\rm t}(t):=\lim_{\Delta\rightarrow0}\Phi_0(t,\Delta)$ 
is the true value for the infinite resolution and $\eta$ is a constant. 
Since a 2nd order finite difference method is partially used in our numerical code, 
we expect at least the 2nd order convergence to our numerical code, that is $n\geq2$. 
As is shown in Fig.~\ref{convergence}, 
the value of $n$ can be read off as $n\sim3.5$ from our numerical results, 
and $\Phi_{\rm t}(t)$ and $\eta$ can be evaluated  
by using the least square fitting assuming $n=3.5$.
\subsection{Threshold} 
First, we summarize our parameter setting for each value of 
the initial radius $r_0$ of the domain wall in Table~\ref{table}.
For $\lambda=1000\mu^{2}$ and $\lambda=2000\mu^{2}$ cases, 
the thresholds $r_{\ast}$ of the BH formation 
are given by
$1.4556243366/\mu$ and  $2.199078357/\mu$, respectively.
\begin{table}
\begin{center}
$\lambda=1000\mu^2$
\\
\begin{tabular}{|c|c|c|c|}
\hline
Sub/Super critical&Initial domain wall radius~[$\mu^{-1}$]&Grid interval~[$\mu^{-1}$]&Param-1/2\\
\hline
Super&$2.500000000\sim 1.455625000$&$5.0\times 10^{-5}$&Param-1\\
\hline
Super&$1.455624950\sim 1.455624600$&$2.5\times 10^{-5}$&Param-1\\
\hline
Super&$1.455624500\sim 1.455624350$&$1.0\times 10^{-6}$&Param-2\\
\hline
\hline
Sub   &$1.455624200\sim 1.455624000$&$1.0\times 10^{-6}$&Param-2\\
\hline
Sub   &$1.455623500\sim 1.440000000$&$5.0\times 10^{-5}$&Param-1\\
\hline
\end{tabular}
\\
\vspace{0.5cm}
$\lambda=2000\mu^2$
\\
\begin{tabular}{|c|c|c|c|}
\hline
Sub/Super critical&Initial domain wall radius~[$\mu^{-1}$]&Grid interval~[$\mu^{-1}$]&Param-1/2\\
\hline
Super&$4.200000000\sim 2.216000000$&$5.0\times 10^{-5}$&Param-1\\
\hline
Super&$2.206000000\sim 2.199162000$&$5.0\times 10^{-5}$&Param-1\\
\hline
Terminated&$2.215000000\sim 2.207000000$&\multicolumn{2}{c|}{Terminated}\\
\hline
Super&$2.199160000\sim 2.199078380$&$2.0\times 10^{-5}$&Param-1\\
\hline
Super&$2.199078377\sim 2.199078367$&$5.0\times 10^{-6}$&Param-2\\
\hline
\hline
Sub   &$2.199078332\sim 2.199077000$&$5.0\times 10^{-6}$&Param-2\\
\hline
Sub   &$2.199076000\sim 2.199010000$&$2.5\times 10^{-5}$&Param-1\\
\hline
Sub   &$2.199000000\sim 2.194000000$&$5.0\times 10^{-5}$&Param-1\\
\hline
\end{tabular}
\end{center}
\caption{
Table of the parameter region 
of our numerical simulation
and the grid interval 
for each simulation.
The leftmost column shows whether the parameter is in the super critical region or 
the subcritical region.
}
\label{table}
\end{table}

In the $\lambda=2000\mu^{2}$ case,
there is 
the parameter region 
in which the lapse function diverges at the origin during the time evolution. 
This behavior might suggest that the 
time slice condition is not appropriate for this parameter region.
However, the value of $r_0$ in this
region is 
far from the threshold value, and it is not a matter to 
investigate the critical behavior.

Apart from the parameter region where the calculation is terminated,
we do not observe the discrete change of the black hole formation time, 
and non trivial phase transition such as 
the phase transition between the delayed collapse phase and 
the prompt collapse phase 
reported in the massive scalar field case~\cite{Okawa:2013jba}. 

\subsection{Mass scaling and the fine structure}
\begin{figure}
\begin{center}
\includegraphics[scale=0.91]{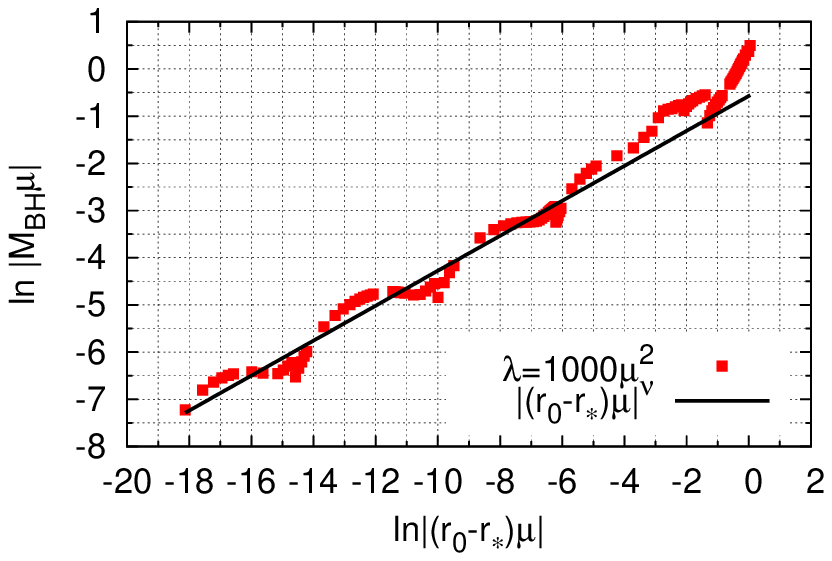}
\includegraphics[scale=0.91]{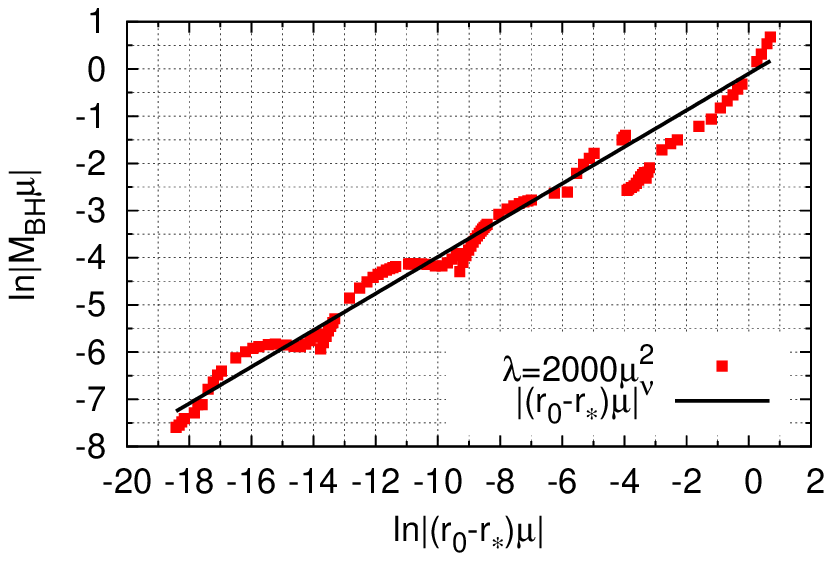}
\caption{
The left panel shows the mass scaling 
for the $\lambda=1000\mu^{2}$ case, 
and the right panel shows the mass scaling 
for the $\lambda=2000\mu^{2}$ case.
In each panel, 
the dots denote the 
numerical results,
and the line denotes the fitting function 
$M_{\mbox{\footnotesize{BH}}}=\zeta  |(r_{0}-r_{\ast})\mu|^{\nu}$.
We determined the coefficient $\zeta$ and $\nu$ by using the least squares fitting 
in the region $\ln |(r_{0}-r_{\ast})\mu|<-5$.
As a result, for $\lambda=1000\mu^{2}$, $\zeta\simeq 0.564$, $\nu\simeq 0.370$ and for $\lambda=2000\mu^{2}$, $\zeta\simeq 0.905$, $\nu\simeq 0.388$.
}
\label{mass_scaling}
\end{center}
\end{figure}

We define
the black hole mass $M_{\mbox{\footnotesize{BH}}}$ 
as the 
half of the apparent horizon radius 
at the moment of 
the apparent horizon formation. 
In the supercritical region($r_{0}>r_{\ast}$),
we can see the mass scaling around the threshold of 
the black hole formation(see Fig.\ref{mass_scaling}).
As we can see in Fig.~\ref{mass_scaling},
the indices of the scaling 
agree with the index of the massless scalar system.

\begin{figure}
\begin{center}
\includegraphics[scale=0.91,clip]{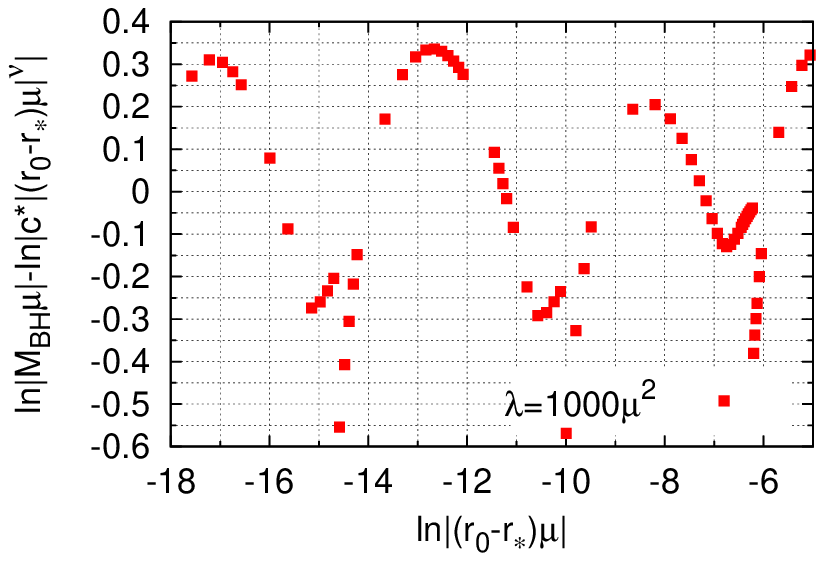}
\includegraphics[scale=0.91,clip]{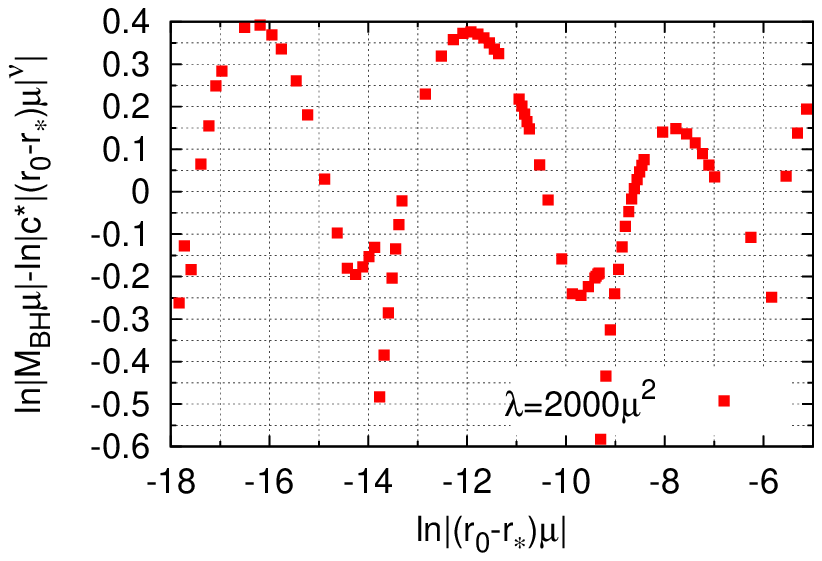}
\caption{
The difference between $M_{\mbox{\footnotesize{BH}}}$ and the 
scaling relation $\zeta|(r_0-r_{\ast})\mu|^\nu$.
The left panel shows the case $\lambda=1000\mu^{2}$, 
and the right panel shows the case $\lambda=2000\mu^{2}$. 
The periodic behaviors
appear in $\ln |(r_0-r_{\ast})\mu|\lesssim-9$.
The periods are about 4.5 in the log scale.
}
\label{fine structure of mass scaling}
\end{center}
\end{figure}

We can also see the fine structure in Fig.~\ref{mass_scaling}.
In order to see the fine structure more clearly, 
we show the difference between $M_{\mbox{\footnotesize{BH}}}$ and the scaling 
relation $\zeta|(r_{0}-r_{\ast})\mu|^\nu$ in 
Fig.~\ref{fine structure of mass scaling}.
The periodic behaviors
appear in the region $\ln |(r_0-r_{\ast})\mu|\lesssim -9$.
The periods 
are about 4.5 in the log scale, which is close to the massless scalar case.

\subsection{Mass discontinuity}
\begin{figure}
\begin{center}
\includegraphics[scale=0.8,clip]{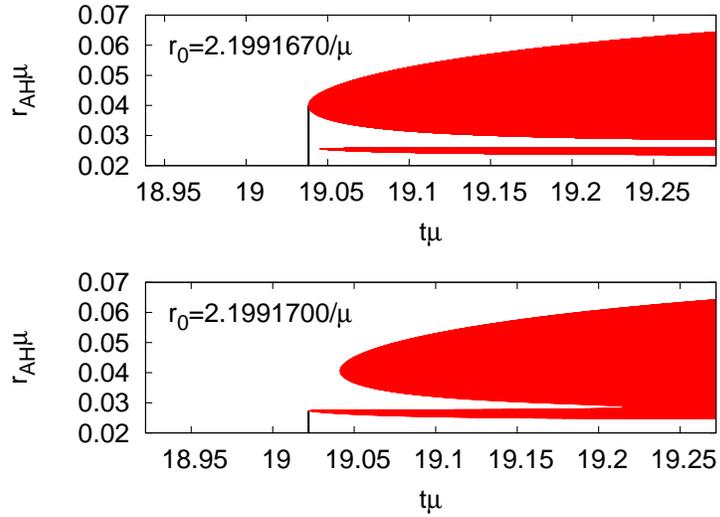}
\caption{
Time evolution of the 
trapped region and marginally outer-trapped surfaces.
The upper panel shows the case $r_{0}=2.1991670/\mu$ and,
the lower panel shows the case $r_{0}=2.1991700/\mu$.
The vertical line on the each graph 
corresponds to the moment of the apparent horizon formation.
}
\label{time evolution of the AH}
\end{center}
\end{figure}

There are some discontinuities 
in the behavior of the mass as a function of $r_0$
($\ln |(r_0-r_{\ast})\mu|\simeq -10$, $-14.5$ for $\lambda=1000\mu^{2}$, 
and $\ln |(r_0-r_{\ast})\mu|\simeq -9.5$, $-14.0$ for $\lambda=2000\mu^{2}$).
The reason for this behavior can be understood by looking at
the time evolution of 
the marginally outer-trapped surfaces, 
whose outgoing null expansion vanishes.
We depict the time evolution of 
the trapped region and 
the marginally outer trapped surfaces
in Fig.~\ref{time evolution of the AH}. 
In Fig.~\ref{time evolution of the AH}, 
the trapped region is described by the shaded region. 
In some period of time, 
the trapped region is divided into two disconnected thick spherical shell regions. 
Therefore, 
we have two connected sequences of outer boundaries of the 
connected trapped regions. 
Consequently, the mass of the black hole at the moment of 
the horizon formation depends on which sequence 
appears first. 
The discrete transition of the black hole mass may happen at the 
parameter for which two sequences simultaneously appear. 
Since the multiple connected trapped regions are essential for this 
phenomenon, 
it cannot be realized in calculations with the areal polar gauge 
or the null coordinates. 
Similar behavior is also reported for the massless scalar system 
with a horizon penetrating gauge condition\cite{Cao:2016tvh}. 

\subsection{Scaling in the subcritical region}
\begin{figure}
\begin{center}
\includegraphics[scale=0.91,clip]{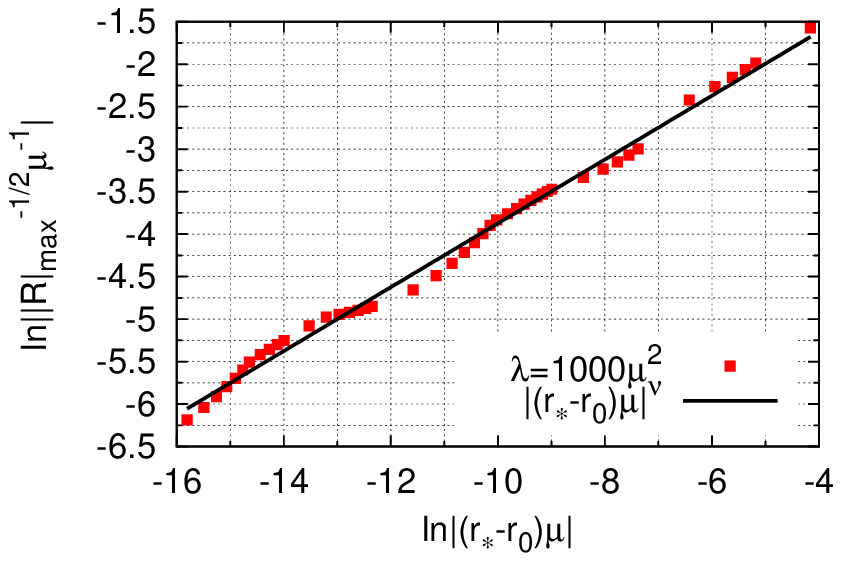}
\includegraphics[scale=0.91,clip]{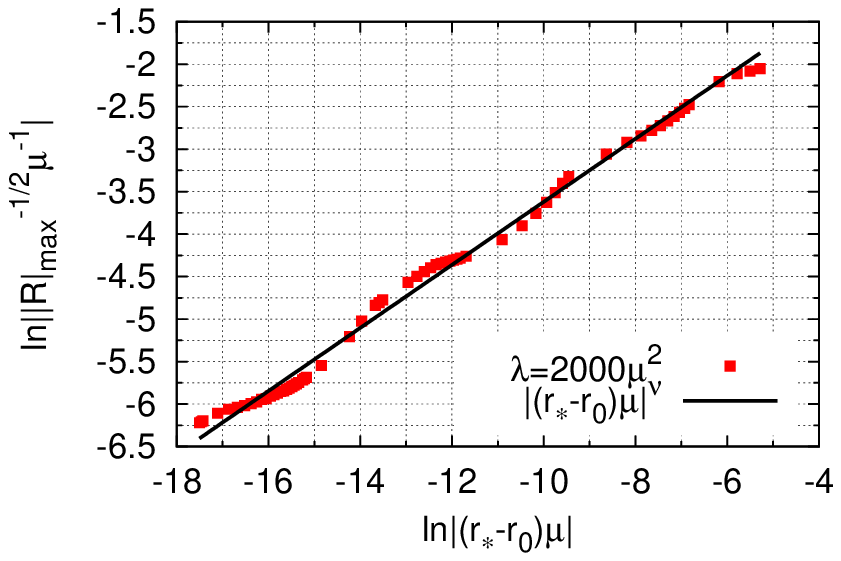}
\caption{
The relation between the maximum value of the curvature and 
the initial radius of the domain wall. 
We plot $|R|_{\mbox{\footnotesize{max}}}^{-1/2}$ as a function of $|(r_{\ast}-r_{0})\mu|$. 
The left panel shows the case $\lambda=1000\mu^{2}$, 
and the right panel shows the case $\lambda=2000\mu^{2}$. 
The black line denotes the fitting function $\zeta|(r_{\ast}-r_{0})\mu|^{\nu}$. 
The coefficient is determined
by the least squares fitting in the region $\ln |(r_{\ast}-r_{0})\mu|<-5$. 
As a result, we obtain $\nu\simeq 0.376$ for $\lambda=1000\mu^{2}$, and $\nu\simeq 0.371$
for $\lambda=2000\mu^{2}$. 
}
\label{curvature scaling}
\end{center}
\end{figure}

Let us consider the scaling behavior in the subcritical parameter region.
We calculate the maximum absolute value of the curvature $R$ and $R_{\mu\nu}R^{\mu\nu}$ 
at the origin in the subcritical parameter region ($r_0<r_{\ast}$), 
which are denoted as $|R|_{\mbox{\footnotesize{max}}}$ and $|R_{\mu\nu}R^{\mu\nu}|_{\mbox{\footnotesize{max}}}$, respectively. 
As is mentioned in the introduction,
it is expected that these curvatures also obey scaling laws. 
Because the behaviors of these curvatures are almost same as each other, 
only the relation between $|R|_{\mbox{\footnotesize{max}}}$ 
and the initial radius of the domain wall is depicted in Fig.~\ref{curvature scaling}.
As is shown in Fig.\ref{curvature scaling}, 
the behavior of the maximum value of the curvature 
agrees with the massless case. 
This is the expected result because, 
if the system comes to be governed by the scaling properties, 
the kinetic energy of the scalar field dominates and the potential term 
becomes less effective near the critical point~\cite{Choptuik:1994ym}. 

\begin{figure}
\begin{center}
\includegraphics[scale=0.91,clip]{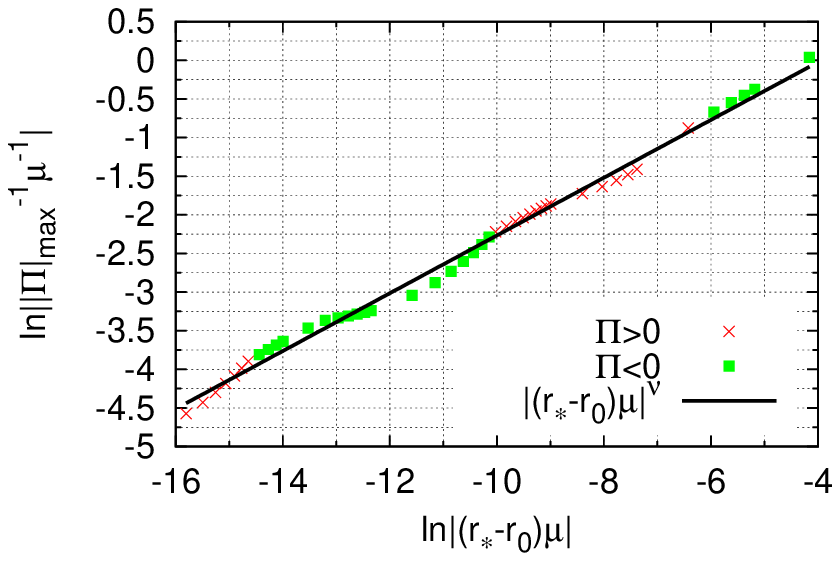}
\includegraphics[scale=0.91,clip]{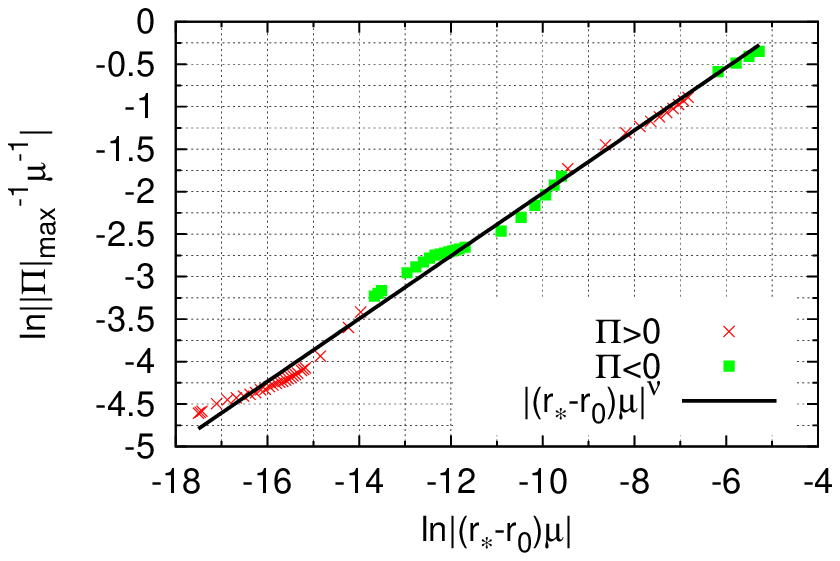}
\caption{
The relation between the absolute maximum value of the conjugate momentum 
of the scalar field and the initial radius of the domain wall.
The left panel shows the case $\lambda=1000\mu^{2}$,
and the right panel shows the case $\lambda=2000\mu^{2}$. 
We plot $|\Pi|_{\mbox{\footnotesize{max}}}$ as a function of $|(r_{\ast}-r_{0})\mu|$. 
Each red(green) point corresponds to a case in which  
the maximum absolute value is realized by a positive (negative) value of 
the conjugate momentum. 
We can fit $|\Pi|_{\mbox{\footnotesize{max}}}$ by the function 
$\zeta|(r_{\ast}-r_{0})\mu|^{-\nu}$ (black line).
By using the least squares fitting,
$\nu$ is
given by $0.374$ for $\lambda=1000\mu^{2}$,
and $0.370$ for $\lambda=2000\mu^{2}$. 
}
\label{pi scaling}
\end{center}
\end{figure}

As is also expected from the above consideration,
we observe the scaling of the absolute maximum value of 
the conjugate momentum $|\Pi|_{\mbox{\footnotesize{max}}}$ 
at the origin in the subcritical region
(see Fig.~\ref{pi scaling}). 
The index is expected from 
the dimensional analysis\cite{Koike:1995jm}, 
because the dimension of the conjugate momentum of 
the scalar field is the inverse of the length scale. 
We can find that the sign of $\Pi$ at the maximum absolute value 
synchronizes with the period of the fine structure. 
It visualizes that there is relation between 
the oscillation of the scalar field near the origin 
and the period of the fine structure in the parameter space. 

\section{summary and discussion}\label{summary}
In this paper, 
we have focused on the spherically symmetric 
minimally coupled scalar field system with a double well potential, 
and investigated
gravitational collapse of 
a spherically symmetric domain wall.
We have 
performed full general relativistic numerical simulation 
for two specific cases which have different parameter sets from 
the previous work\cite{Clough:2016jmh}.
As a result, 
the Choptuik's scaling and the fine structure
have been confirmed. 
We have found that the index of the scaling and the period of the fine structure 
are close to 
the massless scalar case. 
Furthermore, 
for the subcritical region, 
we have checked that the maximum value of the curvature at the origin 
also obeys the scaling law 
with a similar index to the massless scalar case. 

We have shown that the behavior of the black hole mass 
as a function of the initial radius of the domain wall 
is not necessarily smooth. 
Due to the non-trivial structure of the trapped region, 
the behavior of the black hole mass may have discontinuity. 
This behavior is peculiar in the analysis with 
spacelike horizon penetrating time slices.
The origin of this discontinuity is the appearance of multiple connected trapped 
regions.
Recently, similar behavior is reported
in Ref.~\cite{Cao:2016tvh}. 
They found that a new outer horizon appears after 
an apparent horizon initially appears, 
and the difference between the old apparent horizon 
and the new apparent horizon also obeys the scaling law. 
It may be interesting to check if the same scaling low 
can be realized with a non-trivial potential and 
discuss their differences. 
We leave it as a future work.

In this paper, 
the analyses have been done for two specific values 
of the parameter contained in the scalar field potential.
In the supercritical region, 
when the initial radius of the domain wall changes, 
the black hole formation time changes smoothly. 
In this sense,
non-trivial phase transition, such as the transition between 
the prompt collapse phase and the delayed collapse phase 
reported in the massive scalar system~\cite{Okawa:2013jba}, has not been found. 
Apparently, we need further parameter search to complete the phase diagram 
of the spherical domain wall collapse as in the case of 
the massive scalar field~\cite{Okawa:2013jba}.

\section{Acknowledgements}
We would like to thank Hirotada Okawa, Tomohiro Harada, Rong-Gen Cai and Kazuhisa Okamoto for helpful discussion.
This work was 
supported by JSPS KAKENHI Grant Numbers JP16K17688, JP16H01097(C.Y.).

\end{document}